# Sign Reversal and Tunability of Exchange Bias in Nanoscale AuFe Alloy Film: A New Material for Spintronic Application

Compesh Pannu[1*], Santanu Ghosh[1*], Pankaj Srivastava[1], K Bharuth-Ram[2], V. R. Reddy[3], Ajay Gupta[4], Debdulal Kabiraj[5], D. K. Avasthi[4]

[1]Indian Institute of Technology Delhi, New Delhi – 110016
[2]School of Physics, University of Natal, Durban 4041, South Africa
[3]UGC-DAE Consortium for Scientific Research, Indore- 452 017
[4]Amity University, Noida, Uttar Pardesh - 201313
[5]Inter University Accelerator Centre, New Delhi – 110067
[*]compesh@gmail.com, santanu1@physics.iitd.ac.in

**Abstract**

We report here sign reversal and tunability of exchange bias in AuFe cosputtered films of thickness ≈ 63 nm. As deposited film exhibits exchange bias effect at room temperature without external triggering field and its magnitude increases gradually with decrease in temperature down to 5 K. Upon irradiation with 100 MeV $Au^{9+}$ ions at a fluence of $5\times10^{13}$ ions/cm$^2$, hysteresis loop shifts completely from origin towards positive field side at room temperature and reverses sign when temperature is reduced to 5 K as studied by SQUID magnetometry. A well defined uniaxial magnetic anisotropy has been seen by magneto optical Kerr effect (MOKE) in as deposited film as well as in irradiated one. The results are explained on the basis of stress induced magnetic anisotropy in thin films.

**Introduction**

Spintronics aims to exploit the spin degree of freedom of electrons for an advanced generation of electronic devices. The major objective of such an advanced technology is to reduce power consumption while enhancing processing speed, integration density and functionality in comparison with present day complementary semiconductor electronics. Several major new developments, such as the magnetic tunnel junction [1] based on the tunneling magnetoresistance [2], have already lead to significant steps forward. However, commonly used devices, e.g., in today's hard disc drive read heads, or a future magnetic, fast and nonvolatile memory, rely on a ferromagnetic (FM) reference material that needs to be coupled to an antiferromagnetic (AFM) material. The interface coupling of these FM-AFM materials is known as exchange bias effect. A variety of systems such as nanoparticles [3], epitaxial bilayers [4-5] or polycrystalline thin films [6], have shown exchange bias effect. In all of them, exchange bias could be achieved when a FM is field-cooled from a temperature beyond the Neel point of AFM to a temperature much lower than that, or it is fabricated under an external magnetic field. Therefore, in addition to cooling, it is necessary to have an external field to trigger the exchange bias so that a loop shift in the magnetization curve can be observed [7-8]. However, some recently published papers have reported the observation of exchange bias without external magnetic field and referred as spontaneous exchange bias [9-10]. For example, very large spontaneous exchange bias has recently been found in several spin-glass-FM/AF nanocomposite systems under zero-field cooling procedure indicating that exchange bias can be realized without an external trigger field [9]. A truly practical device would be possible in future if instead of zero field cooling procedure, exchange bias could be realized at room temperature or temperature of operation could be raised to room temperature without any external field.

In most of the reports, exchange bias field is found to be either positive or negative, but sign reversal of exchange bias field with tunability is rare. Zhang *et al.* [11] reported the tuning and sign reversal of exchange bias via increasing the maximum measurement field in $Mn_{0.7}Fe_{0.3}NiGe$ alloy. The sign reversal of exchange bias by varying the temperature is reported in $Sm_{0.975}Gd_{0.025}Cu_4Pd$ [12] and $La_{0.2}Ce_{0.8}CrO_3$ [13]. The tuning and sign reversal of exchange bias using external parameters like temperature and ion irradiation has potential applications in thermally assisted random access memory devices [14].

The present study investigates AuFe alloy thin film. Combining the magnetic and optical properties of Au and Fe in a single nanostructure may develop interesting properties for spintronic applications like large magnetic anisotropy [15], large magneto-optical responses [16], high magneto resistance [17] and spin Hall effects [18]. To the best of our knowledge, there are no reports available in literature on AuFe alloy thin film showing exchange bias effect. In the present study, spontaneous exchange bias is observed in AuFe alloy films, even though Au is known to be non magnetic and Fe is ferromagnetic in nature. Some reports have shown evidence of spin polarization transfer between the magnetic Fe and the non-magnetic Au, thereby inducing finite magnetization in Au [19-21]. The phenomenon of development of exchange bias effect in magnetic nanoparticles just by an introduction of non magnetic material is uncommon and its origin is still not understood.

Since exchange bias is an interface phenomenon, introducing defects or disorder at the interfaces modifies the exchange bias properties. Ion irradiation offers unique possibilities to introduce defects and disorder at the interface and consequently modifying exchange bias [22]. Most of the ion induced modification studies on exchange bias systems are found in the low energy regime (eV – keV) [23-24]. Schmalhorst et al. [25] have studied the influence of ion bombardment in exchange bias properties of magnetic tunnel junctions. They observed that by ion bombardment in the presence of a magnetic field, exchange bias can be initiated without field cooling. Effect of low energy ion irradiation in exchange bias is well studied both theoretically and experimentally but studies on modification of exchange bias by high energy ion irradiation are rare.

Au and Fe were cosputtered using RF magnetron sputtering technique to synthesize AuFe thin films. A fraction of $^{57}$Fe was also used during sputtering to make thin films active for Mössbauer spectroscopy. These AuFe films were irradiated with 100 MeV Au$^{9+}$ ions at room temperature at a fluence of $5\times10^{13}$ ions/cm$^2$. Magnetic measurements of these films were performed using superconducting quantum interference device (SQUID), Magneto optical Kerr effect (MOKE), and Mössbauer spectroscopy. Hereafter AuFe film irradiated at a fluence of $5\times10^{13}$ ions/cm$^2$ will be named as 'irradiated film' for convenience.

The room temperature Mössbauer spectra of the as deposited and irradiated film is shown in fig. 1. As deposited and irradiated film spectra were fitted with site distribution and hyperfine field distribution method respectively to get the isomeric shift (IS) and quadrupole splitting (QS).

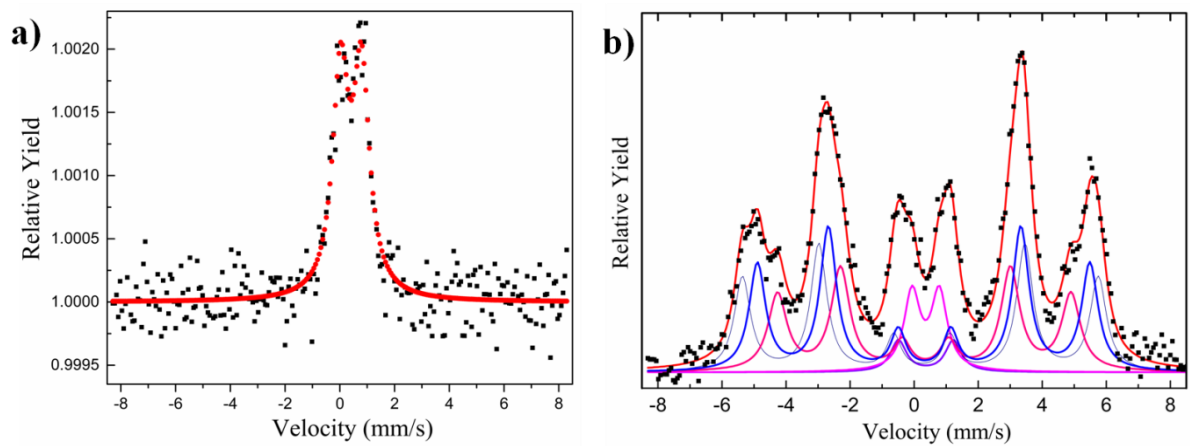

**Fig. 1** Mössbauer spectra of (a) as deposited film, (b) irradiated film.

The as deposited film spectra as shown in fig. 1(a) consists of a doublet with an IS of 0.41 mm/s and a QS of 0.77 mm/s relative to α-Fe. These characteristics show the superparamagnetic behavior of alloy nanoparticles in thin film. These values of IS and QS are consistent with the work of Mishra et al. [26] who reported the Mössbauer spectra of AuFe alloys. The irradiated film spectra as shown in fig. 1(b) consist of three ferromagnetic sextets (B1, B2, and B3) and two paramagnetic doublets (D1, D2). The fit parameters of different components are given in table 1. The hyperfine field of $^{57}Fe$ as reported in literature is 33.5 T [27]. Therefore, the hyperfine field of 33.8 T in the first sextet clearly proves the precipitation of crystallized Fe nanoclusters. The B2 and B3 sextets are ascribed to AuFe alloy nanoparticles in the ferromagnetic state and belongs to iron atoms surrounded by Au atoms. The change in state of some of the alloy nanoparticles in the thin film after irradiation is due to change in size of nanoparticles. After a critical size, alloy nanoparticles show ferromagnetic characteristics. In irradiated film spectra, the line ratios of magnetic components are 3:4:1, which is an evidence that the magnetic moment vector is aligned perpendicular to the angle of emission of $^{57}Fe$ gamma rays i.e. it lies in the plane of AuFe films. The isomer shift (approx 0.3 mm/s) indicates that Fe is in the 3+ charge state. The small contribution from doublets D1 and D2 are due to the presence of superparamagnetic nanoparticles still present in some small regions in thin film. This result is consistent with the observation of Mishra *et al.* [26] who observed such small regions in nanocrystalline AuFe powder. The spectrum shows no evidence of presence of an oxide phase.

Table 1:

| Sample | Peaks | IS (mm/s) | QS (mm/s) | BHF (Tesla) | Ratio L3:L2:L1 | % Area |
|---|---|---|---|---|---|---|
| Pristine | D | 0.41 | 0.77 | - | -- | 100 |
| $5\times10^{13}$ ions/cm$^2$ | D1 | 0.36 | 0.87 | - | | 10 |
| | D2 | 0.37 | 2.09 | - | | 3 |
| | B1 | 0.25 | -0.03 | 33.8 | 3:4:1 | 39 |
| | B2 | 0.38 | 0.02 | 30.3 | 3:4:1 | 41 |
| | B3 | 0.55 | 0.15 | 24.7 | 3:4:1 | 7 |

SQUID measurements were undertaken to obtain the quantitative measure of the irradiation induced changes in the magnetic moment of thin films. As a means of investigating the range and distribution of blocking temperatures, we have measured the moment in a small, constant applied field of 50 Oe, after zero field cooling (ZFC) of the samples well above the maximum blocking temperature. The results of these measurements in as deposited as well as in irradiated film are shown in fig. 2(a). The bifurcation point of FC and ZFC curves (at 57 K) in as deposited and 5e13 film is an indication of maximum blocking temperature, whereas the position of maximum in ZFC curve indicates mean blocking temperature. On irradiation not much changes in maximum blocking temperature was observed. The hysteresis loop at different temperature range for as deposited film is shown in fig. 2(b). The positive shift of hysteresis loop from origin at room temperature shows existence of spontaneous exchange bias in thin film. This shift in hysteresis loop increases from +3.5 Oe to +38.3 Oe as temperature is gradually reduced to 5 K. Note that in the present study no magnetic field was applied during deposition of thin film and no field cooling procedure carried out to induce ordering of spins.

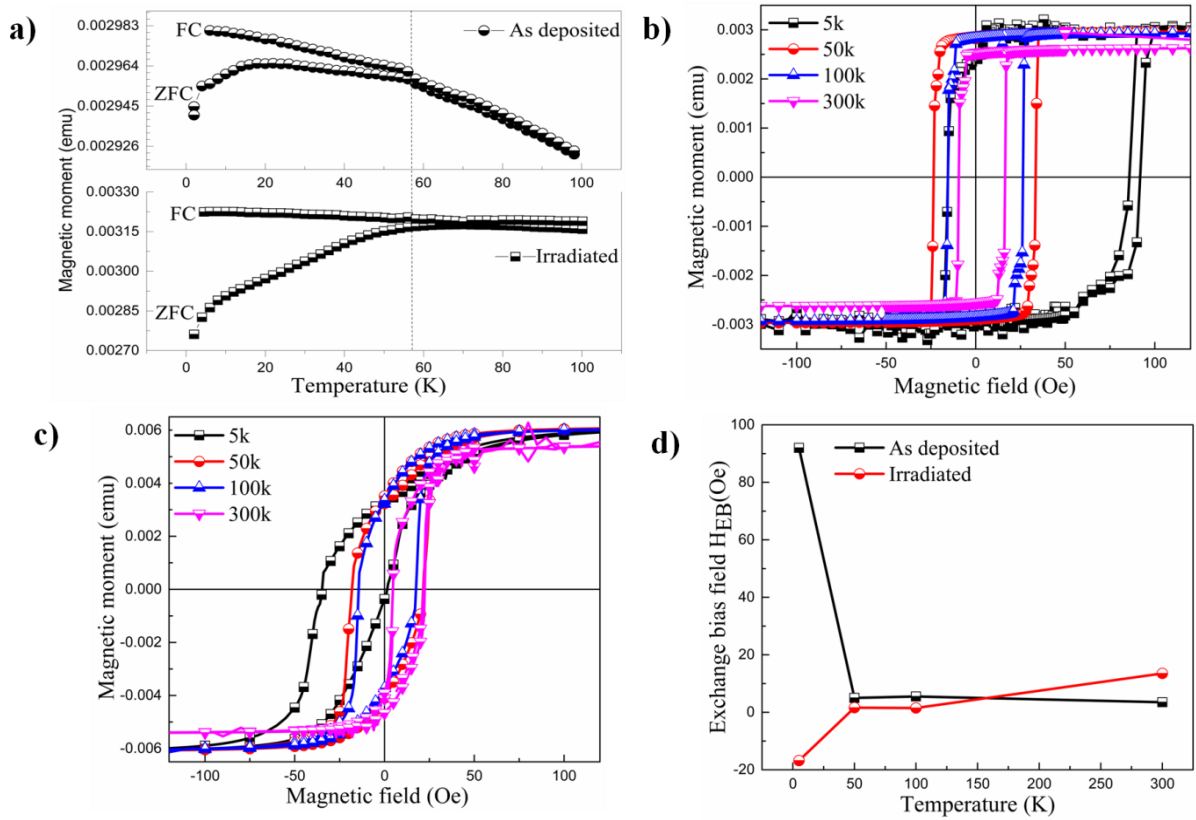

**Fig. 2** a) Temperature dependence of magnetic moment at a small and constant applied field of 50 Oe, b) hysteresis loop of as deposited film at different temperatures, cc) hysteresis loop of 5e13 film at different temperatures, d) temperature dependence of exchange bias field in as deposited and irradiated film.

The observed spontaneous exchange bias in as deposited thin films may be interpreted in terms of interfacial stress induced due to lattice mismatch between Au and Fe. The lattice parameter mismatch develops an interfacial stress during deposition of thin film and this further enhances as the temperature of film is lowered [9]. It has been reported earlier that compacting magnetic nanoparticles under external pressure can result in the development of surface spin disorder [28]. These disordered spins are reported to be highly anisotropic and contributes to rise in magnetic anisotropy [29]. The coupling of these disordered or spin glass like surface spins with core spins of nanoparticles results in exchange bias [29]. This phenomenon is frequently seen in many exchange coupled nanoparticle systems where the pinning layer is a metastable disordered state [30-31].. Such a stress induced magnetic anisotropy can be along easy axis either in positive or negative direction depending on compressive or tensile stress in thin film. Therefore magnetization being aligned along easy axis will be pinned at the interface to the core spins of nanoparticles and this pining gets stronger as the temperature gradually reduced to 5K which consequently results in increase in exchange bias field. The magnetization measurements of irradiated film at different temperatures are shown in fig. 2(c). After irradiation, hysteresis loop completely shifts from origin with exchange bias field of +13.5 Oe at 300 K.

In literature the shift of exchange bias field in opposite direction with changing conditions is known as sign reversal [32] and as shown in fig. 2(d). There is almost no shift in hysteresis loop around maximum blocking temperature which is 57 K. Several groups have reported that exchange bias field approaches zero near $T_B$ in alloy film and value of $T_B$ varies widely with relative elemental concentration in alloys [33-35]. The sign reversal of exchange bias field observed at 5 K in fig. 2(c) could be interpreted as follows. The stress induced magnetic anisotropy field $B_A$ can be expressed as follows [9]:

$$B_A = \frac{-3\lambda_s \sigma}{M_s} \qquad (1)$$

Where $\lambda_s$ is magnetostriction coefficient, $\sigma$ is stress and $M_s$ is saturation magnetization of film. For tensile and compressive stress, $\sigma$ has a positive and negative sign respectively. According to this equation, magnetic anisotropy field changes its direction depending on nature of stress present in thin film. For tensile stress $B_A$ is negative, and for compressive stress, it is positive. In irradiated film, stress most probably changes from tensile to compressive as the temperature is gradually lowered to 5K. With change in stress from tensile to compressive, magnetic anisotropy field changes its direction from negative to positive. Due to change in direction of magnetic anisotropy field in irradiated film, exchange bias field reverses its sign. Therefore, it is possible to tune the exchange bias field in either direction with variation in temperature. Fig. 3(a) and Fig. 3(b) gives the azimuthal angle dependent Kerr hysteresis loops for as deposited and irradiated film respectively.

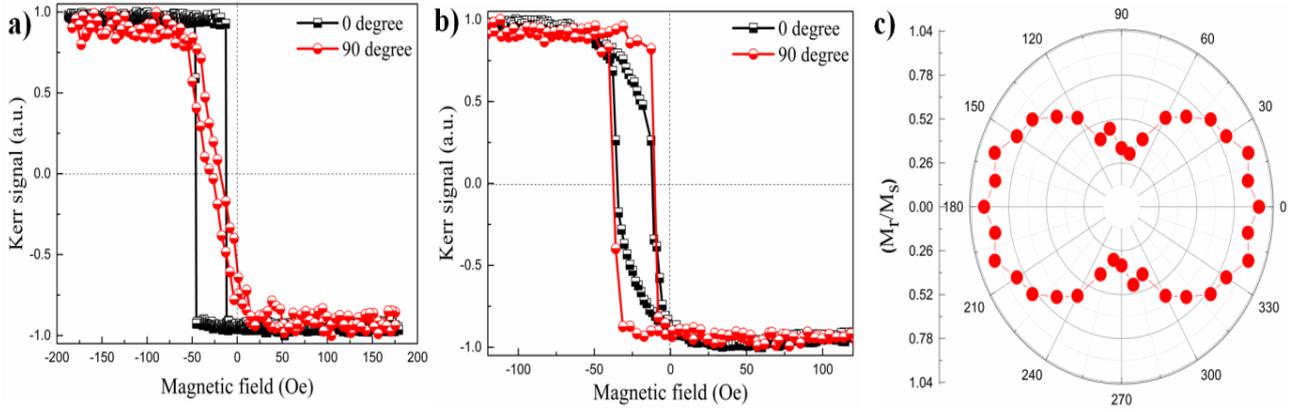

**Fig.3** Hysterisis loops at different azimuthal angles (a) as deposited film, (b) irradiated film, (c) azimuthal angle dependence of remanence magnetization of as deposited film.

The maximum magnetic field applied for MOKE measurements was 200 Oe. In corroboration with SQUID results, MOKE measurements also show shifted hysteresis loops, signature of exchange bias effect. Azimuthal angle dependence of remanence magnetization (polar plot) in as deposited sample is plotted in fig. 3(c). A well defined uniaxial magnetic anisotropy in AuFe film is evident from the polar plot.

**Conclusion**

AuFe cosputtered thin films (≈ 63 nm) prepared by RF magnetron sputtering were irradiated with 100 MeV Au ions. We have observed stress induced spontaneous exchange bias at room temperature which further increases with decrease in temperature of as deposited film. These films have a well defined uniaxial magnetic anisotropy. Further temperature dependent tuning and sign reversal of exchange bias has been observed which make it promising material for MRAM devices, is seen in irradiated AuFe films. Magnetization saturation values in these films are less than 100 Oe which makes them suitable candidate for low field sensor applications.

**Acknowledgement**

We acknowledge financial support from Department of Science and Technology (DST), India, SQUID facility (CRF), Nanoresearch facility, IIT Delhi and accelerator team of IUAC, New Delhi.

**References**

[1] S. Parkin, K. Roche, M. Samant, P. Rice, R. Beyers, R. Scheuerlein, E. O'sullivan, S. Brown, J. Bucchigano, D. Abraham, J. Appl. Phys., **1999**, *85*, 5828.

[2] H. Kumar, S. Ghosh, D. Bürger, L. Li, S. Zhou, D. Kabiraj, D.K. Avasthi, R. Grötzschel, H. Schmidt, J. Appl. Phys., **2011**, *109*, 073914.

[3] D. Sarker, S. Bhattacharya, P. Srivastava, S. Ghosh, Sci. Rep., **2016**, *6*, 39292.

[4] A. Kobrinskii, A. Goldman, M. Varela, S. Pennycook, Phys. Rev. B, **2009**, *79*, 094405.

[5] X. Ke, M.S. Rzchowski, L.J. Belenky, C.-B. Eom, Appl. Phys. Lett., **2004**, *84*, 5458.

[6] K. O'grady, L. Fernandez-Outon, G. Vallejo-Fernandez, J. Magn. Magn. Mater., **2010**, *322*, 883.

[7] J. Nogués, C. Leighton, I.K. Schuller, Phys. Rev. B, **2000**, *61*, 1315.

[8] N. Thuy, N. Tuan, N. Phuoc, N. Nam, T. Hien, N. Hai, J. Magn. Magn. Mater., **2006**, *304*, 41.

[9] N.N. Phuoc, C. Ong, J. Appl. Phys., **2014**, *115*, 143901.

[10] A. Nayak, M. Nicklas, S. Chadov, C. Shekhar, Y. Skourski, J. Winterlik, C. Felser, Phys. Rev. Lett., **2013**, *110*, 127204.

[11] C. Zhang, Y. Mao, R. Wang, H. Xiao, L. Xu, Z. Xia, C. Yang, J. Magn. Magn. Mater., **2017**, *444*, 12.

[12] P. Kulkarni, S. Dhar, A. Provino, P. Manfrinetti, A. Grover, Phys. Rev. B, **2010**, *82*, 144411.

[13] P. Manna, S. Yusuf, R. Shukla, A. Tyagi, Appl. Phys. Lett., **2010**, *96*, 242508.

[14] I. Prejbeanu, M. Kerekes, R.C. Sousa, H. Sibuet, O. Redon, B. Dieny, J. Nozières, Journal of Physics: Condens. Matter, **2007**, *19*, 165218.

[15] V. Gladilin, V. Fomin, J. Devreese, Physica B: Condens. Matter, **2001**, *294*, 302.

[16] Y. Lee, Y. Kudryavtsev, V. Nemoshkalenko, R. Gontarz, J. Rhee, Phys. Rev. B, **2003**, *67*, 104424.

[17] A. Nigam, A. Majumdar, J. Appl. Phys., **1979**, *50*, 1712.

[18] F.W. Fabris, P. Pureur, J. Schaf, V. Vieira, I. Campbell, Phys. Rev. B, **2006**, *74*, 214201.

[19] F. Pineider, C.s. de Julián Fernández, V. Videtta, E. Carlino, A. Al Hourani, F. Wilhelm, A. Rogalev, P.D. Cozzoli, P. Ghigna, C. Sangregorio, ACS nano, **2013**, *7*, 857.

[20] C. de Julián Fernández, G. Mattei, E. Paz, R. Novak, L. Cavigli, L. Bogani, F. Palomares, P. Mazzoldi, A. Caneschi, Nanotechnol., **2010**, *21*, 165701.

[21] F. Wilhelm, P. Poulopoulos, V. Kapaklis, J.-P. Kappler, N. Jaouen, A. Rogalev, A. Yaresko, C. Politis, Phys. Rev. B, **2008**, *77*, 224414.

[22] J. Nogués, J. Sort, V. Langlais, V. Skumryev, S. Surinach, J. Munoz, M. Baró, Phys. Rep., **2005**, *422*, 65.

[23] S. Poppe, J. Fassbender, B. Hillebrands, EPL, **2004**, *66*, 430.

[24] A. Mougin, S. Poppe, J. Fassbender, B. Hillebrands, G. Faini, U. Ebels, M. Jung, D. Engel, A. Ehresmann, H. Schmoranzer, J. Appl. Phys., **2001**, *89*, 6606.

[25] J. Schmalhorst, V. Höink, G. Reiss, D. Engel, D. Junk, A. Schindler, A. Ehresmann, H. J. Appl. Phys., **2003**, *94*, 5556.

[26] A. Mishra, C. Bansal, M. Ghafari, R. Kruk, H. Hahn, Phys. Rev. B, **2010**, *81*, 155452.

[27] M.D. Dyar, D.G. Agresti, M.W. Schaefer, C.A. Grant, E.C. Sklute, Annu. Rev. Earth Planet. Sci., **2006**, *34*, 83.

[28] H. Wang, T. Zhu, K. Zhao, W. Wang, C. Wang, Y. Wang, W. Zhan, Phys. Rev. B, **2004**, *70*, 092409.

[29] S. Chandra, N.F. Huls, M. Phan, S. Srinath, M. Garcia, Y. Lee, C. Wang, S. Sun, O. Iglesias, H. Srikanth, Nanotechnol., **2014**, *25*, 055702.

[30] O. Iglesias, A. Labarta, X. Batlle, Journal Nanosc. Nanotechnol., **2008**, *8*, 2761.

[31] A. Cabot, A.P. Alivisatos, V.F. Puntes, L. Balcells, Ó. Iglesias, A. Labarta, Phys. Rev. B, **2009**, *79*, 094419.

[32] T. Bora, S. Ravi, J. Appl Phys., **2013**, *114*, 183902.

[33] M. Carey, A. Berkowitz, Appl. Phys. Lett., **1992**, *60*, 3060.

[34] M.D. Stiles, R.D. McMichael, Phys. Rev. B, **1999**, *60*, 12950.

[35] S. BRüCK, J. Sort, V. Baltz, S. Surinach, J.S. Munoz, B. Dieny, M.D. Baró, J. Nogues, Adv. Mater., **2005**, *17*, 2978.